\documentclass[sigconf]{aamas}

\usepackage{hyperref}
\usepackage{graphicx}
\usepackage{mathrsfs}
\usepackage[T1]{fontenc}
\usepackage{tikz}
\usetikzlibrary{decorations,decorations.pathmorphing,arrows,shapes,%
  automata,backgrounds,fit,calc,petri,patterns,matrix}
\usepackage{algorithmic}
\usepackage{verbatim}
\usepackage{color}
\usepackage{stmaryrd}
\usepackage{enumerate}
\usepackage{xspace}
\usepackage{balance} 
\usepackage{multirow}
\usepackage{adjustbox}

\usepackage{main}

\setcopyright{ifaamas}
\acmConference[AAMAS '22]{Proc.\@ of the 21st International Conference
on Autonomous Agents and Multiagent Systems (AAMAS 2022)}{May 9--13, 2022}
{Online}{P.~Faliszewski, V.~Mascardi, C.~Pelachaud,
M.E.~Taylor (eds.)}
\copyrightyear{2022}
\acmYear{2022}
\acmDOI{}
\acmPrice{}
\acmISBN{}

\title{Towards Assume-Guarantee Verification of Strategic Ability}
\subtitle{Extended Abstract}

\author{{\L}ukasz~Mikulski$^{1,2}$, Wojciech~Jamroga$^{2,3}$, and Damian Kurpiewski$^{2,1}$}
\affiliation{
  \institution{${}^1$ Faculty of Mathematics and Computer Science, Nicolaus Copernicus University,
Toru{\'n}, Poland}
  \country{${}^2$ Institute of Computer Science, Polish Academy of Sciences, Warsaw, Poland}
}
\affiliation{
  \institution{${}^3$ Interdisciplinary Centre for Security, Reliability and Trust, SnT, University of Luxembourg, Luxembourg}
  \country{\email{lukasz.mikulski@mat.umk.pl, wojciech.jamroga@uni.lu, d.kurpiewski@ipipan.waw.pl}}
  }
\email{lukasz.mikulski@mat.umk.pl,  wojciech.jamroga@uni.lu, d.kurpiewski@ipipan.waw.pl}

\begin{abstract}
Formal verification of strategic abilities is a hard problem.
We propose to use the methodology of assume-guarantee reasoning
in order to facilitate model checking of alternating-time temporal logic with imperfect information and
imperfect recall.
\end{abstract}


\newcommand{\BibTeX}{\rm B\kern-.05em{\sc i\kern-.025em b}\kern-.08em\TeX}
\pgfarrowsdeclare{arrow30}{arrow30}
{
  \arrowsize=0.3pt
  \advance\arrowsize by.275\pgflinewidth%
  \pgfarrowsleftextend{+-\arrowsize}
  \advance\arrowsize by.5\pgflinewidth
  \pgfarrowsrightextend{+\arrowsize}
}
{
  \arrowsize=0.3pt
  \advance\arrowsize by.275\pgflinewidth%
  \pgfsetdash{}{+0pt}
  \pgfsetroundjoin
  \pgfpathmoveto{\pgfqpoint{1\arrowsize}{0\arrowsize}}
  \pgfpathlineto{\pgfqpoint{-12\arrowsize}{2.5\arrowsize}}
  \pgfpathlineto{\pgfqpoint{-12\arrowsize}{-2.5\arrowsize}}
  \pgfpathclose
  \pgfusepathqfillstroke
}

\newcommand{\compos}     {\mathit{Comp}}

\newcommand{\es}         {\varnothing}
\newcommand{\Nat}        {\mathbb{N}}

\newcommand{\ag}[1]{^{(#1)}}

\newcommand{\authorsalt}{{\L}ukasz~Mikulski, Wojciech~Jamroga, and Damian Kurpiewski}

\begin{document}

\pagestyle{fancy}
\fancyhead{}

\maketitle

\section{Introduction}\label{sec:intro}

\emph{Alter\-nating-time temporal logic} \ATLs~\cite{Alur97ATL,Alur02ATL,Schobbens04ATL} and \emph{Strategy Logic} \SL~\cite{Chatterjee10strategylogic,Mogavero14behavioral} provide powerful tools to reason about strategic aspects of MAS.
Specifications in agent logics can be used as input to algorithms and tools for \emph{model checking}~\cite{Alur98mocha-cav,Chen13prismgames,Busard14improving,Huang14symbolic-epist,Cermak14mcheckSL,Lomuscio17mcmas,Cermak15mcmas-sl-one-goal,Belardinelli17publicActions,Belardinelli17broadcasting,Jamroga19fixpApprox-aij,Kurpiewski21stv-demo}.
However, verification of strategic abilities suffers both from state-space and strategy-space explosion.
Even for the more restricted logic \ATL, model checking of its imperfect information variants ranges from \Deltwo--complete to undecidable~\cite{Bulling10verification,Jamroga06atlir-eumas,Schobbens04ATL,Dima11undecidable,Guelev11atl-distrknowldge}.

In this paper, we make the first step towards compositional model checking of strategic properties in asynchronous multi-agent systems with imperfect information and imperfect recall.
To this end, we creatively expand the assume-guarantee framework of~\cite{Lomuscio10assGar,Lomuscio13assGar}.
Instead of searching through the states and strategies of the entire model, we ``factorize'' it
and perform most of the search locally,
using \emph{assume-guarantee reasoning}~\cite{Pnueli84assGuar,Clarke89assGuar}.
We illustrate the approach by means of a simple voting scenario.
Finally, we evaluate the practical gains through verification experiments.

\para{Related work.}
Compositional verification dates back to~\cite{Hoare69axiomatic,Owicki76relyGuar,Jones83relyGuar}, and has been intensively studied for temporal specifications~\cite{Pnueli84assGuar,Clarke89assGuar,Henzinger98assGuar,Devereux03compositionalRA,Kwiatkowska10assGuar,Lomuscio10assGar,Lomuscio13assGar,Fijalkow20assGuar}.
Our approach is based on~\cite{Lomuscio10assGar,Lomuscio13assGar}, where assume-guarantee rules were defined for liveness properties of MAS. A related scheme for \ATL with perfect information strategies and aspect-oriented programs was considered in~\cite{Devereux03compositionalRA}.

\section{Models of Concurrent MAS}\label{sec:models}

We use the MAS representations of~\cite{Lomuscio10assGar,Lomuscio13assGar},
which allow for asynchronous and synchronous composition of local transitions.

\para{Modules.}\label{sec:modules}
Let $D$ be a domain (for all variables used in the system).
For any set of variables $X$, let $D^X$ be a set of all valuation functions on $X$.
We follow~\cite{Lomuscio13assGar}, and
divide the variables in a module into \emph{state variables} and \emph{input variables}.
An agent can read and modify only its state variables.
The input variables are not a part of its state, but their values limit the set of executable transitions.

\begin{definition}[Module~\cite{Lomuscio13assGar}]\label{d:module}
A \emph{module} is $M=(X,I,Q,T,\lambda,q_0)$, where:\
$X, I$ are finite sets of state and input variables, respectively, with $X\cap I=\es$;\
$Q$ is a finite set of states;\
$\lambda:Q\to D^X$ labels each state with a valuation of the state variables;\
$q_0\in Q$ is the initial state;\
$T\subseteq Q\times D^I\times Q$ is a transition relation.
We require that if $(q,\alpha,q')\in T, q\neq q'$, then $(q,\alpha,q)\notin T$,
and for each pair $(q,\alpha)\in Q\times D^I$ there exists $q'\in Q$ such that $(q,\alpha,q')\in T$.
\end{definition}

\begin{figure}[t]\centering
\begin{tabular}{@{}c@{}}
\begin{tikzpicture}[->,auto,>=arrow30,node distance=1.4cm,font=\tiny]
  \node[initial, initial where=below,state] (A)                      {$?,?,?$};
  \node[state]         (B1) [below left of=A, xshift=-30, yshift=15]   {$1,?,?$};
  \node[state]         (B2) [below right of=A, xshift=30, yshift=15]  {$2,?,?$};
  \node[state]         (C1) [below left of=B1]       {$1,1,?$};
  \node[state]         (C2) [below right of=B1]       {$1,!,?$};
  \node[state]         (C3) [below left of=B2]       {$2,!,?$};
  \node[state]         (C4) [below right of=B2]       {$2,2,?$};
  \node[state]         (D1) [below left of=C1, xshift=15]       {$1,1,T$};
  \node[state]         (D2) [below right of=C1, xshift=-15]       {$1,1,F$};
  \node[state]         (D3) [below left of=C2, xshift=15]       {$1,!,F$};
  \node[state]         (D4) [below right of=C2, xshift=-15]       {$1,!,T$};
  \node[state]         (D5) [below left of=C3, xshift=15]       {$2,!,T$};
  \node[state]         (D6) [below right of=C3, xshift=-15]       {$2,!,F$};
  \node[state]         (D7) [below left of=C4, xshift=15]       {$2,2,F$};
  \node[state]         (D8) [below right of=C4, xshift=-15]       {$2,2,T$};
  
  \path (A) edge node [swap] {$\forall$} (B1);
  \path (A) edge node {$\forall$} (B2);
  
  \path (B1) edge node [swap] {$\forall$} (C1);
  \path (B1) edge node {$\forall$} (C2);
  \path (B2) edge node [swap] {$\forall$} (C3);
  \path (B2) edge node {$\forall$} (C4);
  
  \path (C1) edge node [swap] {$T$} (D1);
  \path (C1) edge node {$F$} (D2);
  \path (C2) edge node [swap] {$F$} (D3);
  \path (C2) edge node {$T$} (D4);
  \path (C3) edge node [swap] {$T$} (D5);
  \path (C3) edge node {$F$} (D6);
  \path (C4) edge node [swap] {$F$} (D7);
  \path (C4) edge node {$T$} (D8);
  
  \path (C1) edge [loop right] node [below] {$?$} (C1); 
  \path (C2) edge [loop left] node [below] {$?$} (C2); 
  \path (C3) edge [loop right] node [below] {$?$} (C3); 
  \path (C4) edge [loop left] node [below] {$?$} (C4); 
  
  \path (D1) edge [loop below] node [above] {$\forall$} (D1); 
  \path (D2) edge [loop below] node [above] {$\forall$} (D2); 
  \path (D3) edge [loop below] node [above] {$\forall$} (D3); 
  \path (D4) edge [loop below] node [above] {$\forall$} (D4); 
  \path (D5) edge [loop below] node [above] {$\forall$} (D5); 
  \path (D6) edge [loop below] node [above] {$\forall$} (D6); 
  \path (D7) edge [loop below] node [above] {$\forall$} (D7); 
  \path (D8) edge [loop below] node [above] {$\forall$} (D8); 
  
\end{tikzpicture}
\end{tabular}
\vspace{-0.3cm}
\caption{Module of a voter deciding between two candidates}
\Description{Basic modules of a voter and a coercer used in the leading example.}
\label{fig:voter}
\vspace{-0.3cm}
\end{figure}
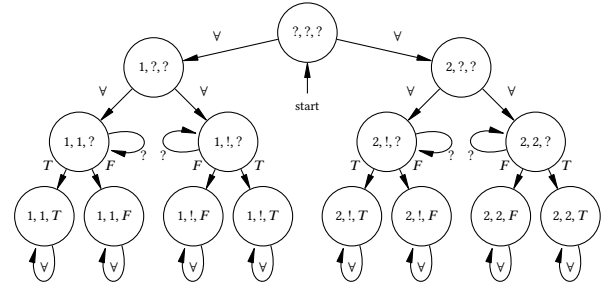

\begin{example}\label{ex1:simple_vote}
We use a voting scenario, inspired by~\cite{Jamroga19fixpApprox-aij,JPSDM20}, consisting of modules $M\ag{1},\dots,M\ag{n}$ of voters and $M\ag{c}$ of a coercer.

Every voter has three local variables in $X\ag{i}$:\
    $vote\ag{i}$: the vote being cast ($?, 1$, or $2$);\
    $reported\ag{i}$: the vote value presented to the coercer ($?, 1, 2$ or $!$), where $!$ means that the voter decided not to show her vote;\
    $pstatus\ag{i}$: the punishment status ($?, T$ or $F$).
Moreover, $I\ag{i}$ consists of variable $pun\ag{i}$ controlled by the coercer.
The voter first casts her vote, then decides whether to share its value with the coercer.
Finally, she waits for the coercer's decision to punish her or to refrain from punishment.
The module is shown in Figure~\ref{fig:voter}.

The coercer has two available actions per voter: to punish the voter or to refrain from punishment. He can
execute them in any order, but only after the respective voters decide to share or not.
\end{example}

Modules $M,M'$ are \emph{asynchronous} if $X\cap X'=\es$.
Note that all the modules presented in Example~\ref{ex1:simple_vote} are asynchronous.

\para{Composition of Modules.}\label{sec:composition}
The model of a MAS is given by the asynchronous composition $M = M\ag{1} | \dots | M\ag{n}$ that combines modules $M\ag{1}, \dots, M\ag{n}$ into a single module $M$~\cite{Lomuscio13assGar}.
The composition is standard; it only requires the compliance of the valuations.

\para{Traces and Words.}\label{sec:traces}
A trace of a module $M$ is an infinite sequence of alternating states and transitions $\sigma=q_0\alpha_0 q_1\alpha_1\ldots$, where
$q_0$ is the initial state and
$(q_i,\alpha_i,q_{i+1})\in T$ for every $i\in\Nat$.
An infinite word $w=v_0 v_1,\ldots \in(D^X)^\omega$ is \emph{derived} by module $M$ with trace $\sigma=q_0\alpha_0 q_1\alpha_1\ldots$ if $v_i = \lambda(q_i)$ for all $i\in\Nat$.
An infinite word $u=\alpha_0 \alpha_1,\ldots \in(D^I)^\omega$ is \emph{admitted} by $M$ with $\sigma$ if $\sigma=q_0\alpha_0 q_1\alpha_1\ldots$ .
Finally, $w$ (resp.~$u$) is derived (resp.~admitted) by $M$ if there exists a trace of $M$ that derives (resp.~admits) it.

\section{What Agents Can Achieve}\label{sec:logic}

\emph{Alternating-time temporal logic} \ATLs~\cite{Alur97ATL,Alur02ATL,Schobbens04ATL}
introduces \emph{strategic modalities} $\coop{C}\gamma$, expressing that coalition $C$ can enforce the temporal property $\gamma$.
In this paper, we use the \emph{imperfect information/imperfect recall} variant without next step operator $\Next$ and nested strategic modalities,
denoted \sATLs (``simple \ATLs'').

\para{Syntax.}\label{sec:atl-syntax}
Formally, the syntax of $\sATLs$ is defined by:
\begin{center}
$\phi ::= p(Y) \mid \neg\phi \mid \phi\land\phi \mid \coop{C} \gamma; \quad
\gamma ::= p(Y) \mid \neg\gamma \mid \gamma\land\gamma \mid \gamma \Until \gamma$.
\end{center}
where $p:Y\to D$ for some subset of domain variables $Y\subseteq X$. That is, each atomic statement refers to the valuation of a subset of variables used in the system.
$\Until$ is the ``strong until'' operator of \LTL.
The ``sometime'' and ``always'' operators $\Sometm$ and $\Always$ can be defined as usual by
$\Sometm\gamma \equiv \top \Until \gamma$ and $\Always\gamma \equiv \neg(\top \Until \neg\gamma)$.

\para{Semantics.}\label{sec:atl-semantics}
A \emph{memoryless imperfect information strategy} for agent $i$
is a function $s_i:Q_i\to T_i$.
We say that a trace $\sigma$ (word derived with $\sigma$) \emph{implements} a strategy $s_i$
if for any $j$ where $q_j^{(i)}\neq q_{j+1}^{(i)}$
we have $s_i(q_j^{(i)})=(q_j^{(i)},\alpha_j,q_{j+1}^{(i)})$, where
$\alpha_j:I_i\to D$ and $\alpha_j(x)=\lambda(q_j)(x)$.

Let $C\subseteq \{1,\ldots,n\}$ be a set of agent indices.
We define \emph{joint strategies} for $C$ as tuples of individual strategies, one per $i\in C$.
The semantics of strategic operators is given by the following clause:
\begin{description2}
\item[{$M,q \satisf \coop{C} \gamma$}] if there exists a joint strategy $s_C$ for $C$ such that, for any word $w$ that implements $s_C$, we have $\model,w \satisf \gamma$.
\end{description2}

\section{Assumptions and Guarantees}\label{sec:assumptions}

We propose an assume-guarantee scheme, where one can reduce the complexity of model checking \sATLs by
verifying individual strategic abilities of single agents against overapproximating abstractions of its environment, i.e., the rest of the system.
The general idea is that if an agent has a successful strategy in a more nondeterministic environment, then it can can use the same strategy to succeed in the original model.
Moreover, it often suffices to prepare the abstraction based only of the modules that are connected with the agent by at most $k$ synchronization steps.

\para{Assumptions and Guarantees.}
The environmental abstractions are formalized by \emph{assumptions} $A=(M_A,F)$,
where $M_A$ is a module and $F$ is a set of accepting states that provide B\"{u}chi-style accepting rules for infinite traces derived by $M$.
The assumption should be constructed so that it \emph{guarantees} that the set of computations accepted by $A$ covers the sequences of changes in the input variables $I_M$ of module $M$.
We capture those changes by the notion of \emph{curtailment}.
Formally, a sequence $v=v_1v_2\ldots$ over $D^Y$ is a curtailment of sequence $u=u_1u_2\ldots$ over $D^X$ (where $Y\subseteq X$)
if there exists an infinite sequence of indices $j_1<j_2<...$ with $j_1=1$
such that $\forall_i \forall_{j_i\leq k<j_{i+1}} v_i=u_k|_Y$.

\para{The Scheme.}
Let $M = M_1 | M_2 | \dots | M_n$ be a system composed from modules $M_1, M_2,\dots, M_n$, where $X_{M_i}\cap X_{M_j}=\es$ for $i\neq j$.
By $\compos_i^1$ we denote the composition of all modules directly related to $M_i$.
Moreover, $\compos_i^k$ denotes the composition of the modules in $\compos_i^{k-1}$ and the modules directly related to them (except for $M_i$).
Further, let $\psi_i, i\in C$ be path formulas of \sATLs, one for each agent in $C$.
Simple assume-guarantee reasoning for strategic ability is provided by the following inference rule:
\[
\bf{R_k}\;\;\;
\begin{array}{c}
\forall_{i\in C}\; M_i | A_i \,\models_{ir}\, \coop{i} \psi_i\\
\forall_{i\in C}\; \compos_i^k \,\models\, A_i\\
\hline
M_1 | ... | M_n \models_{ir} \coop{C} \bigwedge_{i\in C}\psi_i
\end{array}
\]

\section{Experiments}\label{sec:experiments}

\begin{table}[t]
  \begin{adjustbox}{width=1.05\columnwidth,center}
  \begin{tabular}{|c|c|c|c|c|c|c|c|c|}
  \hline
  \multirow{2}{*}{\textbf{V}}  & \multicolumn{4}{c|}{\textbf{Monolithic model checking}}       & \multicolumn{4}{c|}{\textbf{Assume-guarantee verification}}   \\ \cline{2-9}
                                  & \textbf{\#st} & \textbf{\#tr} & \textbf{DFS} & \textbf{Apprx} & \textbf{\#st} & \textbf{\#tr} & \textbf{DFS} & \textbf{Apprx} \\ \hline
  2                           & 529           & 2216          & <0.1         & <0.1/Yes           & 161           & 528           & <0.1         & <0.1/Yes       \\ \hline
  3                           & 1.22e4        & 1.28e5        & <0.1         & 0.8/Yes            & 1127          & 7830          & <0.1         & <0.1/Yes        \\ \hline
  4                           & 2.79e5        & 6.73e6        & <0.1         & 30/Yes             & 7889          & 1.08e5        & <0.1         & 0.5/Yes           \\ \hline
  5                           & 6.43e6        & 3.42e8        & \multicolumn{2}{c|}{timeout}      & 5.52e4        & 1.45e6        & <0.1         & 8/Yes          \\ \hline
  6                           & \multicolumn{4}{c|}{timeout}                                      & 3.86e5        & 1.92e7        & <0.1         & 135/Yes            \\ \hline
  7                           & \multicolumn{4}{c|}{timeout}                                      & \multicolumn{4}{c|}{timeout}                                   \\ \hline
  \end{tabular}
  \end{adjustbox}
\vspace{0.05cm}
\caption{Results of assume-guarantee verification for simple voting (times given in seconds; timeout=2h)}
\label{tab:res}
\vspace{-0.95cm}
\end{table}

Here, we present preliminary experimental results for the assume-guarantee rule proposed in Section~\ref{sec:assumptions},
using the voting scenario of Example~\ref{ex1:simple_vote} as the benchmark.
The assumptions are provided by a simplified module of the coercer, where he only waits
for the value reported by $\mathit{Voter}_1$, no matter how he reacts to other voters' choices.
The algorithms have been implemented in Python, and run on a server with 2.40 GHz Intel Xeon Platinum 8260 CPU, 991 GB RAM, and 64-bit Linux.

The verified formula was
$\varphi \equiv \coop{\mathit{Voter}_1}\Always(\lnot \prop{pstatus_1} \lor \prop{voted_1=1})$.
The results are presented in Table~\ref{tab:res}.
The first column describes the configuration of the benchmark, i.e., the number of the voters.
Then, we report the performance of model checking algorithms that operate on the explicit model of the whole system vs.~assume-guarantee verification.
\emph{DFS} is a straightforward implementation of depth-first strategy synthesis. \emph{Apprx} refers to the method of fixpoint-approximation~\cite{Jamroga19fixpApprox-aij}; besides the time, we also report if the approximation was conclusive.

\section{Conclusion}\label{sec:conlusions}

In this paper, we sketch how assume-guarantee reasoning can be extended for verification of strategic abilities.
The main idea is to factorize coalitional abilities by the abilities of the coalition members, and to verify the individual abilities against B\"{u}chi-style abstractions of the agents' environment of action.
Preliminary experimental evaluation has produced very promising results, showing noticeable improvement in the verification of large models consisting of asynchronous agents with independent goals.

\begin{acks}
We acknowledge the support of the National Centre for Research and Development, Poland (NCBR), and the Luxembourg National Research Fund (FNR), under the PolLux/FNR-CORE project STV (POLLUX-VII/1/2019 -- C18/IS/12685695/IS/STV/Ryan).
\end{acks}

\balance
\bibliographystyle{plain}

\end{document}